\documentclass[10pt,conference]{IEEEtran}
\IEEEoverridecommandlockouts
\usepackage{graphicx}
\usepackage{textcomp}
\usepackage{booktabs}
\usepackage{colortbl}
\usepackage[table]{xcolor}
\usepackage{pifont}
\usepackage{microtype}
\usepackage{xspace}
\usepackage[most]{tcolorbox}
\usepackage{enumitem}
\usepackage{subcaption}
\usepackage{wrapfig}
\usepackage{pgf,pgfmath,etoolbox}
\usepackage{multirow}
\usepackage{tabularx} 
\usepackage{amsmath,amssymb,amsfonts}
\usepackage{algorithmic}
\usepackage{array}
\usepackage{url}
\AtBeginDocument{%
  \providecommand\BibTeX{{%
    Bib\TeX}}}

\usepackage{amssymb}

\providecommand{\cmark}{\ensuremath{\checkmark}}
\providecommand{\xmark}{\ensuremath{\times}}
\providecommand{\pmark}{\ensuremath{\triangle}}
    
\newcommand{\ourbench}{\textsc{R2ABench}}

\def\BibTeX{{\rm B\kern-.05em{\sc i\kern-.025em b}\kern-.08em
    T\kern-.1667em\lower.7ex\hbox{E}\kern-.125emX}}

\definecolor{FindingBgBlue}{RGB}{235, 245, 255}
\definecolor{FindingFrameBlue}{RGB}{100, 150, 200}

\begin{document}

\title{Benchmarking Requirement-to-Architecture Generation with Hybrid Evaluation}

\author{
\IEEEauthorblockN{Minxiao Li, Li Zhang, Shuying Yan, Fang Liu, Liehao Li, Yang Liu, and Xiaoli Lian}
\IEEEauthorblockA{
\textit{State Key Laboratory of Complex \& Critical Software Environment} \\
\textit{School of Computer Science and Engineering} \\
Beihang University, Beijing, China \\
liminxiao@buaa.edu.cn, fangliu@buaa.edu.cn
}
}

\maketitle

\begin{abstract}
Software architecture serves as the blueprint of a software system, capturing high-level structural decisions that shape downstream implementation and system quality. 
Despite this central role, generating architecture designs from requirement documents remains underexplored, with limited task-specific benchmarks for generation or rigorous evaluation.
To bridge this gap, we introduce \ourbench{}, a benchmark for requirements-to-architecture (R2A) reasoning. \ourbench{} contains 68 projects with human-validated references collected from both educational-style settings and GitHub repositories. Each project is packaged as a unified instance that provides a 
structured software requirements specification (SRS), a reference architecture view in PlantUML and the corresponding source requirement artifacts.
To evaluate generated architecture views, we design a layered hybrid evaluation framework spanning syntax validation, structural graph diagnostics, and semantic and evidence-based architecture scoring. 
Using this framework, we conduct a comprehensive empirical study of state-of-the-art LLMs and agents on \ourbench{}. We find that current systems can often generate syntactically valid and readable architecture views, but still struggle with relation-level architecture modeling across both subsets: component identification is substantially stronger than edge recovery, and edge hallucination is the dominant structural failure mode. Semantically, structural fidelity does not guarantee requirement coverage or traceability, which emerge as the primary quality gaps.
The associated data files are available at \url{https://figshare.com/s/01f0a5fb6243a6a60f23}.
\end{abstract}


\section{Introduction}

Software architecture is a critical bridge between requirements and implementation \cite{nuseibeh2001weaving, medvidovic2010software}. It describes the core components of a system, their responsibilities, hierarchical relationships, deployment logic, and interactions with external systems. Architecture views are essential artifacts for expressing and communicating such architectural decisions \cite{IEEE42010}. As defined by Rozanski and Woods \cite{rozanski2012software}, a view is a representation of one or more structural aspects of an architecture that explains how the architecture addresses stakeholder concerns. Through architecture views, complex systems can be decomposed into understandable representations for different concerns, supporting communication and collaboration among developers, architects, operators, and business stakeholders \cite{clements2003documenting, tekinerdogan2011defining}.

Despite their importance, deriving a faithful architecture view from requirements remains a costly and inconsistent activity. It is far from a mechanical translation: an analyst must reconcile functional requirements with non-functional requirements, technical constraints, domain rules, and architecturally significant requirements (ASRs), and synthesize them into a coherent structure of components, boundaries, dependencies, data stores, and external interfaces. Because this synthesis is performed manually, the requirements-to-architecture mapping is often produced inconsistently and with weak traceability between design decisions and the requirements that justify them. The resulting views are also hard to keep faithful: an analysis of 15,000 architecture views across 12,200 projects finds that roughly 75\% are never updated after creation \cite{migliorini2024architectural}. This makes requirements-to-architecture reasoning a natural target for automated support.

Large language models (LLMs) offer a new opportunity to address this long-standing problem. 
In recent years, generative AI has been increasingly adopted in code generation, testing, program repair, and software maintenance, and recent studies have begun to explore the use of LLMs to support architectural activities, ranging from design-decision support and architecture analysis to runtime architectural adaptation~\cite{esposito2025generative}. However, as noted by Esposito et al.~\cite{esposito2025generative}, the capability of LLMs to generate architecture views remains largely underexplored. This task is not a straightforward extension of code generation. 
LLMs are strong at producing code snippets and completing local functions, but these are largely local, self-contained tasks. Architecture views instead demand system-level reasoning---understanding broad stakeholder concerns, selecting appropriate abstraction levels, recognizing architectural patterns, and reflecting quality-attribute requirements---so whether the coding strengths of LLMs carry over to this setting is far from obvious. 
Similarly, general-purpose coding agents with task decomposition, tool use, and memory have been applied to complex software engineering tasks, but it remains unclear whether they can reliably reason from requirements to architecture views.

Answering this question empirically calls for a benchmark: requirements paired with reference architecture views, together with a way to score a generated view against a reference. Existing resources do not meet this need.
Prior work mostly targets text-to-model generation for individual artifact types from curated descriptions~\cite{alaswad2026plantucd,shbita2025mermaidseqbench,chen2023automated}, while architecture-recovery resources extract views from existing source code rather than generating them from requirements~\cite{migliorini2024architectural,sathvika2026llm}. Studies that do generate architecture from requirements are typically evaluated on a few case studies, without a reusable benchmark or reference views to compare against~\cite{tagliaferro2025leveraging,helmi2025arlo}. 
Evaluation is a further open problem: architecture generation admits no single executable oracle, unlike repository-level coding benchmarks where executable validation provides a success signal~\cite{jimenez2023swe}, so text-based metrics and single-judge LLM checks are both inadequate~\cite{papineni2002bleu,zheng2023judging}. We compare existing benchmarks in detail in Section~\ref{sec:related-work}. The absence of both a dedicated benchmark and a reliable evaluation methodology jointly hinders progress in this area.
 
To address both gaps, we introduce \ourbench{}, a benchmark for requirements-to-architecture (R2A) reasoning. Specifically, \ourbench{} contains 68 projects with human-validated references drawn from educational settings and GitHub repositories. Each instance is organized around a shared SRS schema that normalizes requirements into functional requirements, non-functional requirements, technical constraints, domain rules, actors, data objects, external systems, and ASRs. Beyond this specification, each project provides a reference architecture view in PlantUML that can be parsed into a typed component graph, the corresponding source requirement artifacts, and requirement-to-evidence trace links. Building on this benchmark, and to overcome the evaluation difficulties above, we design a layered hybrid evaluation framework that combines syntax validation, structural graph diagnostics, semantic and evidence-driven architecture scoring, and human calibration, so that architecture quality is assessed through complementary signals rather than any single, biased metric.

Using \ourbench{}, our empirical study of state-of-the-art LLMs and agents shows that current systems generate syntactically valid and readable views but remain limited at relation-level modeling, identifying components far more reliably than they recover dependencies.

Our contributions are as follows:
\begin{itemize}[leftmargin = *,topsep=0pt]
    \item We introduce \ourbench{}, a benchmark for requirements-to-architecture reasoning, containing 68 projects with human-validated references from educational courses and GitHub repositories.

    \item We design a layered hybrid evaluation framework that combines syntax validation, structural graph diagnostics, semantic and evidence-driven architecture scoring.

    \item We conduct an extensive empirical study of state-of-the-art LLMs and agents on \ourbench{}, evaluating architecture generation across structural fidelity, semantic adequacy, and evidence traceability.
\end{itemize}

\section{Task Formulation}

We frame R2A as a forward synthesis task: given a project $p$, the generator $G$ derives an architecture view $\hat{\mathcal{V}}_p$ from a structured software requirements specification $\mathcal{S}_p$ alone, rather than recovering one from existing code:
\begin{equation}
    \hat{\mathcal{V}}_p = G(\mathcal{S}_p).
\end{equation}
For evaluation, each project is additionally annotated with a reference architecture view $\mathcal{V}_p^{*}$.

The input $\mathcal{S}_p$ is a normalized software requirements specification. To provide a standardized requirements representation, we structure the SRS with reference to ISO/IEC/IEEE 29148~\cite{iso2018iso}:
\begin{equation}
    \mathcal{S}_p =
    \langle FR, NFR, TC, DR, A, D, ES, ASR \rangle,
\end{equation}
where $FR$ denotes functional requirements, $NFR$ non-functional requirements, $TC$ technical constraints, $DR$ domain rules, $A$ actors, $D$ data objects, $ES$ external systems, and $ASR$ architecturally significant requirements. This representation provides a common task interface across projects, while retaining the functional, quality-attribute, and constraint information relevant to architectural decisions.

The output $\hat{\mathcal{V}}_p$ is an architecture view, represented as a typed component graph that captures the system's structural decomposition. Each view specifies the major architectural elements and their responsibilities, the relations among them, layers or boundaries, external systems, and key data stores. Concretely, we serialize $\hat{\mathcal{V}}_p$ as PlantUML code. We adopt PlantUML rather than image-based diagrams because its textual form can be generated directly by LLMs and, unlike free-form images, can be deterministically rendered for inspection and parsed into a graph for automatic evaluation.

\section{Data Construction}

\subsection{Overview}

Building a benchmark for R2A reasoning faces a fundamental tension: evaluation requires clean, well-structured references with unambiguous ground truth, whereas software developed in practice rarely provides them. To address this, \ourbench{} pairs two complementary subsets. 
The educational subset (E-R2A) consists of graduate-level course projects whose requirements documents follow the ISO/IEC/IEEE 29148 standard and whose reference views were peer-reviewed in course settings; these properties make E-R2A well-suited for rigorous and reproducible evaluation. The subset of projects collected from GitHub (G-R2A), in contrast, consists of openly developed software whose architectures are more tightly coupled and whose requirements are not documented in any single, explicit specification. Together, the two subsets let \ourbench{} assess R2A capability under both idealized and repository-grounded conditions.

\begin{figure}[t]
    \centering
    \setlength{\abovecaptionskip}{0.1cm}
    \includegraphics[width=\linewidth]{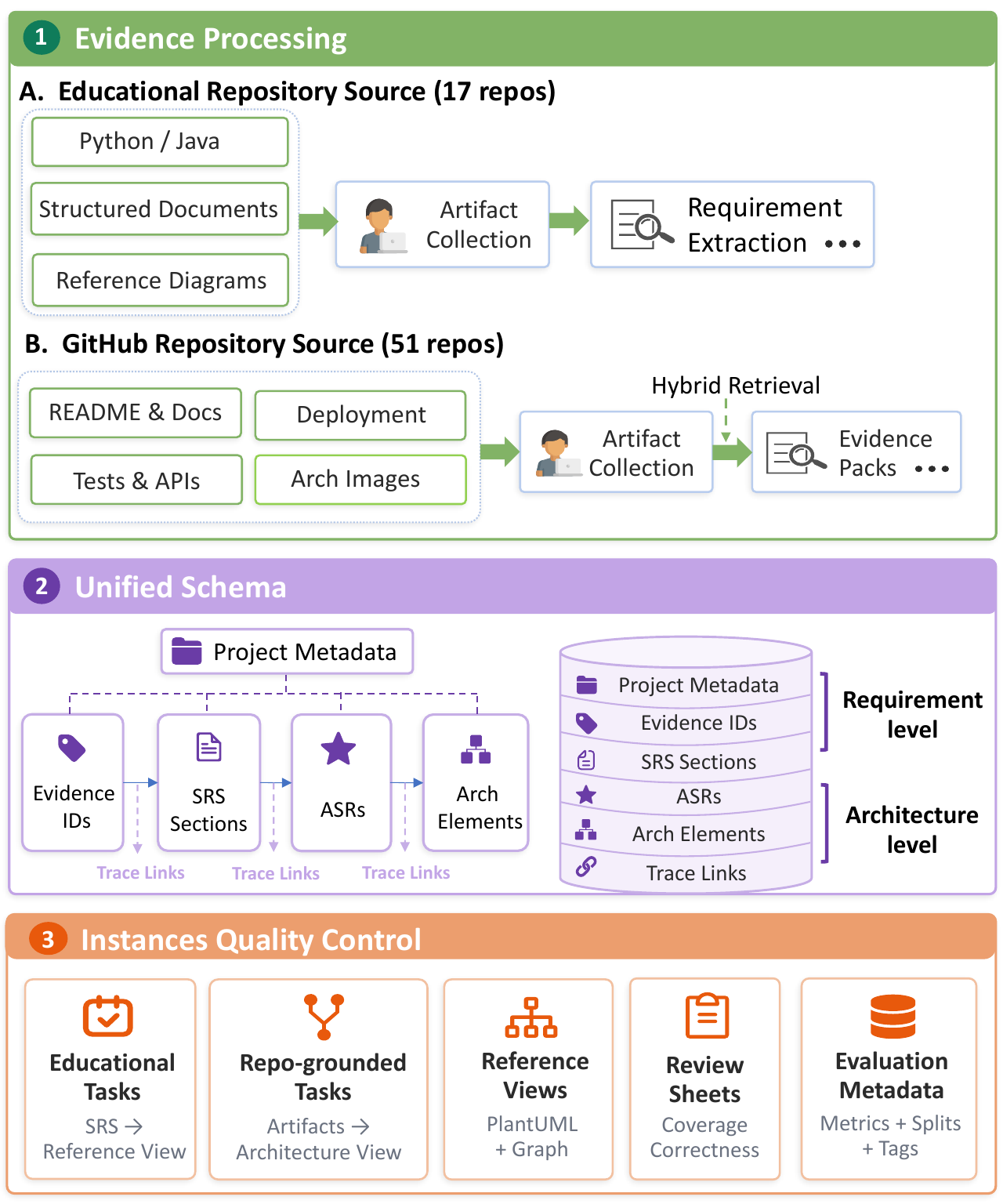}
    \caption{Overview of the \ourbench{} construction pipeline: evidence processing, unified schema, and quality control.}
    \label{fig:benchmark}
    \vspace{-15pt}
\end{figure}

\subsection{E-R2A: Educational-Project Benchmark Subset}

E-R2A is the educational-project subset in \ourbench{}, targeting R2A scenarios where project architectures are relatively well-structured and requirements documents are comparatively clear. The data are collected from graduate-level software engineering course projects. Each project contains a software requirements analysis document, an architecture view, and the corresponding source requirement artifacts. The requirements documents were authored following the ISO/IEC/IEEE~29148 requirements-engineering standard~\cite{iso2018iso}, and both the documents and the diagrams were reviewed and graded as part of the course process, providing vetted reference artifacts for evaluation.

To ensure the selected projects are suitable for architecture view generation and evaluation, we apply three criteria: \ding{182} the target programming language must account for more than 50\% of the total code by bytes; \ding{183} the system must contain a complete and clear architecture view to serve as the reference artifact; and \ding{184} the system must maintain a clear front-end/back-end separation and a reasonably modular structure. Following these criteria, E-R2A contains 17 project samples.

The construction of E-R2A consists of three steps: requirements extraction, architecture view standardization, and reference artifact validation. First, we use rule-based extraction to convert the standard-compliant course documents into the SRS format defined in this study, and we attach evidence trace links that map each extracted requirement back to its source location in the original document. Second, we use a model to convert the original architecture views into PlantUML representations, so that the reference architectures can be parsed and compared by the evaluation scripts. Two annotators then cross-validate each converted view against the original diagram to ensure structural consistency in component hierarchy, layer assignment, component relations, and overall structure. The corresponding extraction rules, prompts, and processing scripts are available in the project repository.

\subsection{G-R2A: GitHub Repository Subset}

G-R2A is designed to reflect architecture reasoning conditions in open-source projects. Compared with E-R2A, projects collected from GitHub usually involve more complex system structures, richer technology stacks, longer evolution histories, and more diverse architectural representations. We use 373 manually coded architecture-view samples and their corresponding repositories from prior work~\cite{migliorini2024architectural} as the candidate pool, 
and retain repositories with sufficient supporting materials and parsable architecture views through a candidate scoring and manual inspection procedure, which yields 51 GitHub repository samples in the current version.

\textbf{Candidate scoring and selection.}
We score each candidate repository with a deterministic, rule-based procedure that computes scores from repository metadata and content statistics along six dimensions: documentation scale, evidence-category coverage, requirement relevance, temporal consistency, traceability potential, and SRS convertibility. Candidates are retained based on preset thresholds on the total and per-dimension scores; the complete rubric, thresholds, and computation rules are released in our project repository for reproducibility. This stage initially selects 52 repositories; after manual inspection, one is removed, yielding 51 valid G-R2A samples.

\textbf{Evidence collection.}
For each retained repository, we then assemble the source material from which its SRS will be reconstructed. 
We first resolve a source-code snapshot temporally aligned with its manually coded architecture view~\cite{migliorini2024architectural} when metadata is available, mitigating architecture drift caused by repository evolution.
From this snapshot we collect the repository artifacts---which we refer to as \emph{evidence}---spanning documentation (e.g., README, docs, requirements-like files) and implementation (e.g., interface definitions, entry points, configuration). Code artifacts are used to recover architecture-relevant interfaces, constraints, and deployment signals when documentation alone is insufficient.

\textbf{Evidence pack construction.}
Unlike E-R2A, where requirements come from a single standardized document, a GitHub repository has no such specification: architecture-relevant information is sparse and scattered across many heterogeneous files. G-R2A therefore retrieves a small, traceable subset of evidence before generating the SRS.

The retrieval stage first cleans the collected files by removing noisy elements such as markup and links, and splits the cleaned text into evidence chunks. It then issues one query per SRS concern (e.g., functional requirements, non-functional requirements, constraints) over the target SRS schema, and combines three complementary signals over the same chunks: lexical matching via BM25~\cite{robertson2009probabilistic}, vector-space matching via token-frequency cosine similarity with synonym expansion~\cite{salton1975vector}, and code-aware retrieval~\cite{bajracharya2006sourcerer} that prioritizes architecturally relevant files (e.g., interface files for interface-related sections, configuration files for constraint-related ones). The three ranked lists are merged with reciprocal-rank fusion (RRF)~\cite{cormack2009reciprocal}:
\[
\mathrm{RRF}_s(c)=
\sum_{r \in \mathcal{T}}
\frac{w_r}{k+\mathrm{rank}_{r,s}(c)},
\]
where $\mathcal{T}$ is the set of three retrieval signals, $\mathrm{rank}_{r,s}(c)$ is the rank of chunk $c$ under signal $r$ for section $s$, $w_r$ is the fusion weight of signal $r$, and $k$ is a smoothing constant. 
We use heuristic weights ($w_{\mathrm{BM25}}{=}1.0$, $w_{\mathrm{vsm}}{=}0.9$, $w_{\mathrm{code}}{=}0.8$; $k{=}60$) that prioritize exact lexical matching, since requirement and architecture terms often appear verbatim in documentation and identifiers; full settings are in the repository.

\textbf{SRS generation and review.}
We use GPT-5.4 to generate an initial SRS from each evidence pack. The generation constraints enforce traceability and suppress hallucination: the SRS may use only facts from the evidence pack, every requirement must cite at least one evidence id, and unsupported requirements, assumptions, or design content are prohibited; claims inferred from code or configuration are kept distinguishable from those stated in documentation. A separate model, Claude Opus 4.8, then performs an automated review that flags candidate issues over a predefined set of issue types (e.g., unsupported claims, contradictions, missing requirements, traceability gaps).

To control SRS quality, the model-flagged issues undergo human review. Three reviewers, each with about four years of software development experience and familiarity with requirements and architecture documentation, first align on the criteria, then independently label each issue as \texttt{ACCEPT}, \texttt{PARTIAL\_ACCEPT}, or \texttt{REJECT}. We report pairwise agreement and Fleiss' $\kappa$~\cite{fleiss1971measuring} (Table~\ref{tab:srs-human-review}); final decisions follow majority vote, with full disagreements resolved by discussion and recorded as adjudicated decisions. The resulting $\kappa=0.76$ indicates substantial agreement. The SRS is then revised according to the final human decisions, producing the finalized specification used for architecture-view generation.
\begin{table}[t]
\centering
\caption{Human review results for SRS quality control.}
\label{tab:srs-human-review}
\footnotesize
\setlength{\tabcolsep}{3.5pt}
\renewcommand{\arraystretch}{1.08}
\begin{tabular}{@{}lrrr|lr@{}}
\toprule
\multicolumn{4}{c}{\textbf{Decision Distribution}} &
\multicolumn{2}{c}{\textbf{Agreement Summary}} \\
\cmidrule(lr){1-4}\cmidrule(l){5-6}
\textbf{Reviewer} & \textbf{A} & \textbf{PA} & \textbf{R} &
\textbf{Metric} & \textbf{Value} \\
\midrule
R1 & 257 & 37 & 46 & Full / majority agreement & 288 / 52 \\
R2 & 247 & 47 & 46 & Avg. pair agreement & 0.8980 \\
R3 & 241 & 51 & 48 & Fleiss' $\kappa$ & 0.7630 \\
\midrule
\multicolumn{6}{@{}l}{\scriptsize A=ACCEPT, PA=PARTIAL\_ACCEPT, R=REJECT.} \\
\bottomrule
\end{tabular}
\vspace{-8pt}
\end{table}

\textbf{Reference view standardization.}
The steps above produce the SRS input for each repository; we now describe how its reference architecture view is prepared. Each retained repository has a manually coded architecture view from prior work~\cite{migliorini2024architectural}. Following the same procedure as E-R2A, a model converts each view into parsable PlantUML and two annotators cross-validate it against the original for consistency in component hierarchy, layer assignment, relations, and overall structure, yielding the reference views used for L1 evaluation. We treat each coded view as an authored reference at its recorded timestamp, to which the source-code snapshot and evidence are aligned, so comparisons reflect the architecture as documented rather than the latest repository state.

\textbf{Benchmark statistics.} 
Table~\ref{tab:benchmark-stats} reports four structural properties of the reference views---\emph{components} (architectural elements), \emph{relations} (dependencies), their ratio (\emph{relations per component}, i.e., coupling density), and \emph{layers} (hierarchical depth)---characterizing how large, interconnected, and deeply structured each reference architecture is.
E-R2A projects provide cleaner source requirements and more standardized reference views, but their reference views contain more components and deeper layer structures (reflecting an explicitly modularized design). G-R2A projects, by contrast, often have fewer components but substantially denser relations (relations per component of 1.00 vs.\ 0.38). This relational density, rather than the number of components, is what characterizes the structural complexity of projects collected from GitHub.

\begin{figure}[t] 
\includegraphics[width=\columnwidth]{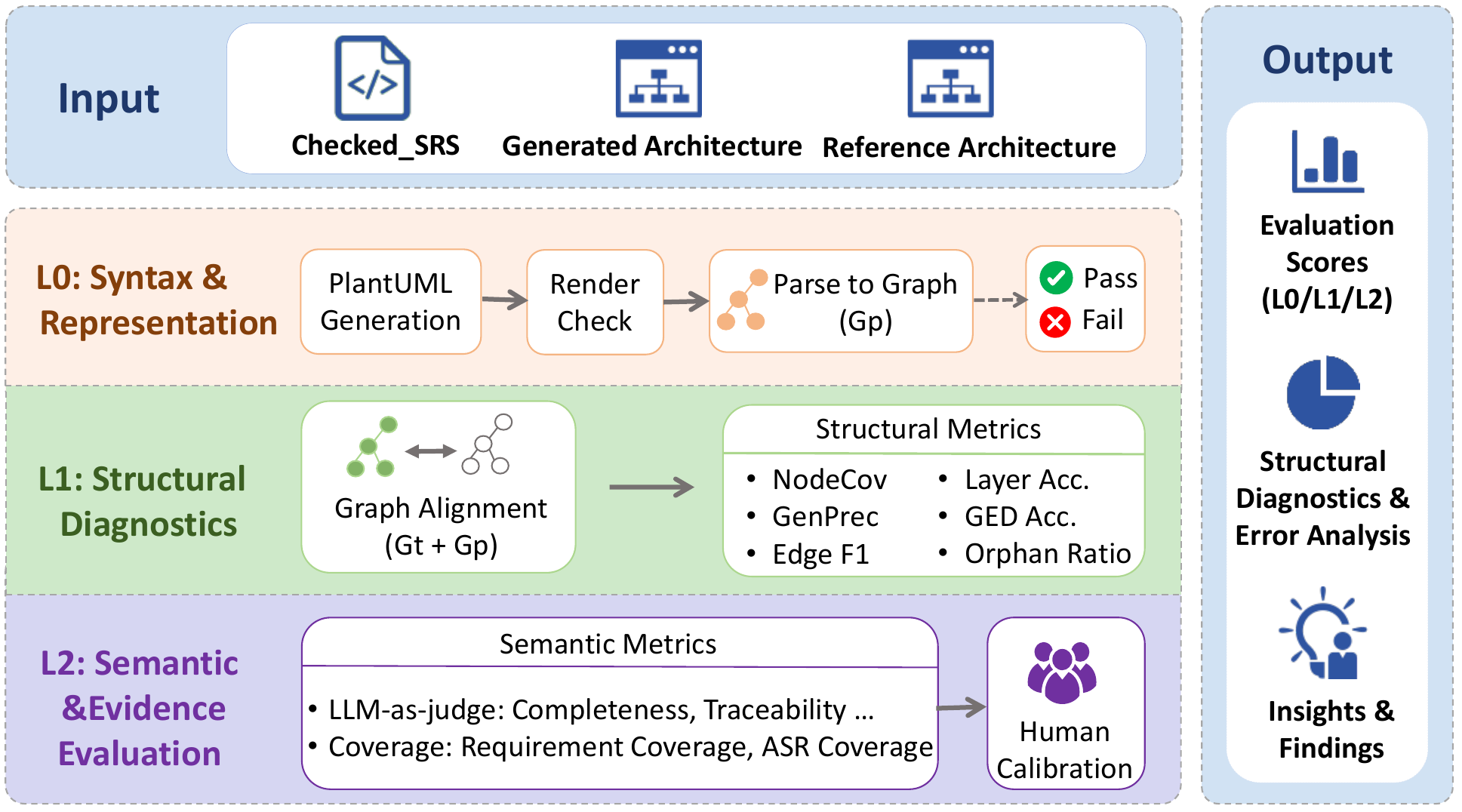}
    \caption{Overview of the layered hybrid evaluation framework, comprising L0 syntax validation, L1 structural diagnostics, and L2 semantic and evidence evaluation.}
    \label{fig:evaluation}
    \vspace{-10pt}
\end{figure}

\begin{table}[t]
\centering
\caption{Statistics of reference architecture views in \ourbench{}.}
\label{tab:benchmark-stats}
\begin{tabular}{lccc}
\toprule
 & Educational & GitHub & Overall \\
\midrule
\ Projects                 & 17 & 51 & 68 \\
\ Components (avg.)         & 22.9 & 13.5 & 15.9 \\
\ Relations (avg.)          & 8.8 & 13.5 & 12.4 \\
\ Relations per component  & 0.38 & 1.00 & 0.78 \\
\ Layers (avg.)             & 8.0 & 3.5 & 4.6 \\
\bottomrule
\end{tabular}
\vspace{-15pt}
\end{table}

\section{Evaluation Framework}

\subsection{Overview}

We adopt a layered hybrid evaluation framework for the generated artifacts. Unlike code generation, architecture views cannot be validated through executable tests or \texttt{pass@k} metrics \cite{chen2021evaluating}: the same requirements may admit multiple reasonable representations, with no single executable oracle or uniquely correct diagram \cite{IEEE42010}.
We evaluate the generated artifacts at three levels: L0 syntax and representation validity, L1 structural diagnostics, and L2 semantic and evidence-based quality—complemented by human calibration on a sampled subset.

At L0, we check whether the generated PlantUML artifact, which serializes the generated architecture view $\hat{\mathcal{V}}_p$, can be rendered successfully and parsed into a directed architecture graph $G_p=(V_p,E_p)$.
At L1, when a reference view $\mathcal{V}_p^{*}$ is available, we parse it into a reference graph $G_r=(V_r,E_r)$ and compare $G_p$ against $G_r$ using graph-based metrics such as node matching, edge matching, layer accuracy, and graph edit distance. 

At L2, we use an LLM-as-a-Judge evaluator~\cite{zheng2023judging} to rate each view $\hat{\mathcal{V}}_p$ along five quality dimensions (Table~\ref{tab:l2-dimensions}).


\subsection{L0: Syntax and Representation Validity}

\textbf{L0} acts as a gate: it renders the PlantUML output and parses it into nodes and edges, and samples that fail either step are marked failed and excluded from all downstream graph-based metrics.

\subsection{L1: Structural Diagnostics}

L1 evaluates the structural similarity between the generated architecture view and the reference view. The evaluator parses both into directed architecture graphs $G_p=(V_p,E_p)$ and $G_r=(V_r,E_r)$, obtained from the generated view $\hat{\mathcal{V}}_p$ and the reference view $\mathcal{V}_p^{*}$, respectively. Here, $V$ represents architecture elements, such as services, components, databases, and external systems, while $E$ represents directed dependencies or data-flow relations. We further write $V_{L,p}\subseteq V_p$ and $V_{L,r}\subseteq V_r$ for the leaf (component-level) nodes of each graph. Each parsed node retains its layer path, such as \texttt{Backend::Auth Service}, enabling the evaluator to check not only whether a component is generated, but also whether it is placed in the correct architectural layer.

Because LLM-generated views may use component names or abstraction levels that differ from those in the reference view, L1 first uses GPT-5.4-mini to perform semantic alignment between generated nodes and reference nodes before computing graph metrics.
Since internal nodes serve primarily as hierarchical grouping containers, alignment is restricted to leaf nodes, which represent the indivisible system components that carry architectural meaning. The alignment result is represented as:
\begin{equation}
    M=\{(R_j,P_j,b_j)\}_{j=1}^{m},
\end{equation}
where $R_j \subseteq V_{L,r}$ denotes one or more reference leaf nodes, $P_j \subseteq V_{L,p}$ denotes one or more generated leaf nodes, and $b_j \in \{0,1\}$ indicates whether the matched group is placed in a semantically correct layer.

\begin{figure}[t]
    \centering
    \setlength{\abovecaptionskip}{0.1cm}
    \includegraphics[width=\linewidth]{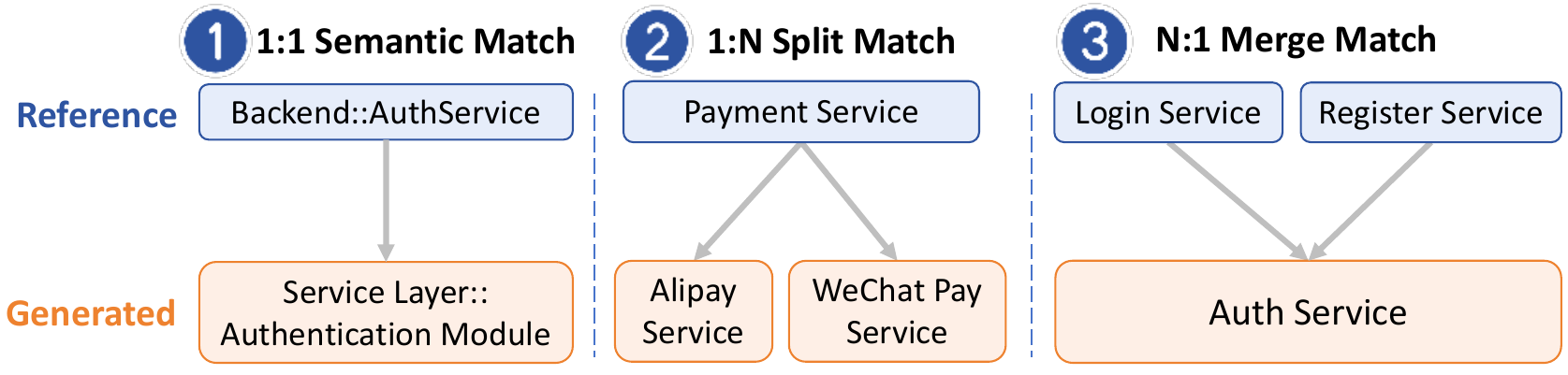}
    \caption{Three forms of semantic correspondence under granularity mismatch: one-to-one, split, and merge between reference and generated components.}
    \label{fig:alignment}
    \vspace{-15pt}
\end{figure}

Because the generated and reference views need not decompose the system at the same granularity, a reference component may map to a generated one as one-to-one, split (one reference realized by several generated components), or merge (several reference components collapsed into one), as illustrated in Fig.~\ref{fig:alignment}. The matched groups in $M$ are then used to compute node coverage, generation precision, edge F1, layer accuracy, GED-based accuracy, and graph-structure anomaly metrics.

\begin{enumerate}[leftmargin=*]
    \item \textbf{Node Coverage and Generation Precision ($NodeCov$, $GenPrec$):}
    Because a generated view may realize the reference at a different abstraction level, we score nodes as a granularity-tolerant correspondence, read from the alignment $M$ in two directions. A reference leaf $v\in V_{L,r}$ is covered ($c(v)=1$) if it appears in some matched group of $M$, i.e.\ at least one generated node realizes its responsibility, and $c(v)=0$ if it is unmatched
    (a missing component). Symmetrically, a generated leaf $u\in V_{L,p}$ is grounded ($g(u)=1$) if it appears in some matched group, i.e.\ it is supported by a reference concept, and $g(u)=0$ if it is unmatched (a hallucinated component). We report
    \begin{equation}
    \mathrm{NodeCov}=\frac{1}{|V_{L,r}|}\sum_{v\in V_{L,r}} c(v),
    \end{equation}
    \begin{equation}
    \mathrm{GenPrec}=\frac{1}{|V_{L,p}|}\sum_{u\in V_{L,p}} g(u).
    \end{equation}
    $NodeCov$ and $GenPrec$ thus expose missing and hallucinated components respectively. Because a coarse generated node may cover several reference leaves and a finer set may jointly cover one, we report the two directional scores separately rather than as a single F1, keeping omission and over-generation distinguishable.

    \item \textbf{Edge Correctness ($F1_{edge}$):}
    This metric evaluates whether the model correctly generates invocation or data-flow relations between components. A predicted edge is counted as correct if its source and target nodes correspond to the source and target nodes of a reference edge under the alignment $M$. Edge precision, recall, and F1 are defined as:
    \begin{equation}
    P_{edge}=\frac{TP_E}{TP_E+FP_E}, \quad
    R_{edge}=\frac{TP_E}{TP_E+FN_E},
    \end{equation}
    \begin{equation}
    F1_{edge}=\frac{2P_{edge}R_{edge}}{P_{edge}+R_{edge}}.
    \end{equation}
    Here, $TP_E$ denotes matched edges, $FP_E$ denotes unmatched generated edges, and $FN_E$ denotes unmatched reference edges.

    \item \textbf{Layer Accuracy ($Layer_{Acc}$):}
    This metric evaluates whether matched components are placed in the correct architectural layer:
    \begin{equation}
    Layer_{Acc}=\frac{1}{|M|}\sum_{j=1}^{|M|} b_j,
    \end{equation}
    where $b_j=1$ if the $j$-th matched group is placed in a correct architectural layer and $b_j=0$ otherwise.

    \item \textbf{Normalized GED Score ($GED_{acc}$):}
    This metric evaluates the overall topological similarity between $G_p$ and $G_r$. To reduce the effect of graph size, we normalize the raw GED into an accuracy score from $0$ to $100$. Let $s_p=|V_p|+|E_p|$ and $s_r=|V_r|+|E_r|$:
    \begin{equation}
    GED_{acc}=100\cdot \max\left(0,\,
    1-\frac{GED(G_p,G_r)}{\max(s_p,s_r)}\right).
    \end{equation}
    Here, $GED(G_p,G_r)$ denotes the graph edit distance between $G_p$ and $G_r$; a higher score indicates greater topological similarity.

    \item \textbf{Orphan Ratio ($R_{orphan}$):}
    This metric measures the proportion of generated components with no incoming or outgoing edges:
    \begin{equation}
    R_{orphan}=\frac{|\{v\in V_{L,p}:\deg(v)=0\}|}{|V_{L,p}|}.
    \end{equation}

\end{enumerate}

\subsection{L2: Semantic and Evidence Evaluation}

L2 is the primary layer for evaluating architecture quality. It assesses whether the generated architecture is semantically reasonable, requirement-aware, and supported by available evidence. Specifically, L2 adopts an LLM-as-a-Judge evaluation strategy~\cite{zheng2023judging}, implemented with GPT-5.4-mini, and rates each generated view $\hat{\mathcal{V}}_p$ along five dimensions, 
organized with reference to established software architecture evaluation and documentation-quality criteria~\cite{IEEE42010,iso2018iso,clements2003documenting,ISO25010}: completeness, faithfulness, architectural rationality, readability, and traceability. These dimensions are summarized in Table~\ref{tab:l2-dimensions}.

Beyond these judge-rated dimensions, we additionally compute two judge-based coverage proxies that quantify how much of the requirement space is reflected in the generated view. Given a set of normalized requirements $\mathcal{R}$ and a generated view $\hat{\mathcal{V}}_p$, requirement coverage is defined as:
\begin{equation}
ReqCov=\frac{|\{r\in\mathcal{R}: covered(r,\hat{\mathcal{V}}_p)=1\}|}{|\mathcal{R}|}\times 100\%,
\end{equation}
where $covered(r,\hat{\mathcal{V}}_p)=1$ if the LLM judge determines that requirement $r$ is addressed by at least one component or relation in the view. Similarly, for architecturally significant requirements $\mathcal{R}_{ASR}$:
\begin{equation}
ASRCov=\frac{|\{r\in\mathcal{R}_{ASR}: covered(r,\hat{\mathcal{V}}_p)=1\}|}{|\mathcal{R}_{ASR}|}\times 100\%.
\end{equation}

\begin{table}[t]
\centering
\small
\caption{L2 semantic and evidence-based evaluation dimensions.}
\label{tab:l2-dimensions}
\begin{tabular}{p{0.23\columnwidth}p{0.67\columnwidth}}
\toprule
\textbf{Dimension} & \textbf{Evaluation Focus} \\
\midrule
Completeness & Whether the view covers key functional requirements, ASRs, external systems, data stores, constraints, and quality drivers. \\
\midrule
Faithfulness & Whether generated components, relations, technologies, and assumptions are supported by source evidence or justified inference. \\
\midrule
Architectural Rationality & Whether the decomposition, dependencies, responsibilities, boundaries, and quality-attribute handling are architecturally plausible. \\
\midrule
Traceability & Whether architecture elements and decisions can be linked to requirements, ASRs, constraints, or evidence snippets. \\
\midrule
Readability & Whether the view is understandable at the intended abstraction level, with clear naming, organization, boundaries, and relation labels. \\
\bottomrule
\end{tabular}
\vspace{-10pt}
\end{table}

\begin{table}[t]
  \centering
  \setlength{\abovecaptionskip}{0.1cm}
  \caption{LLMs and agent frameworks for RQ1.}
  \label{tab:models_agents}
  \renewcommand{\arraystretch}{1.2}

  \begin{tabular*}{\columnwidth}{@{\hspace{0.5em}\extracolsep{\fill}}ll@{}}
    \toprule
    \textbf{   Category} & \textbf{Model} \\
    \midrule
    \multirow{2}{*}{   Open-source LLM}
    & DeepSeek V3.2 \\
    & Qwen3-Coder 480B-A35B \\
    \midrule
    \multirow{2}{*}{   Closed-source LLM}
    & Claude Sonnet 4.6 \\
    & GPT-5 \\
    \midrule
    \multirow{3}{*}{   Agent Framework}
    & MetaGPT \cite{hong2024metagpt} \\
    & OpenHands \cite{openhands} \\
    & Mini-SWE \cite{yang2024sweagent} \\
    \bottomrule
  \end{tabular*}

  \vspace{-0.5cm}
\end{table}


Model-based judges are scalable but may be biased, so we calibrate the L2
evaluator against human judgments. The same three reviewers from the SRS review independently score a random sample of 272 views---drawn from the 866 that pass L0 and spanning all model--framework configurations (95\% confidence, 5\% margin of error)---on the five L2 dimensions through a blind static review interface\footnote{Review
UI: \url{https://evalstudio-blind-review-272.netlify.app/}}, following criteria agreed in advance. The L2 judge shows moderate alignment with the resulting human consensus, with 72.77\% of scores within one point and a mean absolute error of 0.53 on the five-point scale.

\section{Experiments}

\subsection{Experimental Setup}

Our empirical study is organized around three research questions:
\begin{itemize}[leftmargin=*,topsep=0pt]
    \item \textbf{RQ1 (Overall capability and input source).} How well do LLMs
    and agents generate architecture views, and how does this differ across
    input sources (educational vs.\ repository-derived)?
    \item \textbf{RQ2 (View characteristics).} How do the behavior and
    granularity of the target view affect generation quality on
    repository-derived projects?
    \item \textbf{RQ3 (Failure modes).} What recurring failure modes arise in
    requirements-to-architecture generation, in terms of structural errors,
    ASR coverage, NFR reasoning, traceability, and unsupported inference?
\end{itemize}

We evaluate a representative set of open-source models, closed-source models, and agentic frameworks (Table~\ref{tab:models_agents}), pairing each framework with each model. MetaGPT is adapted as an agentic baseline for the R2A task---an \texttt{SRSAnalyst}$\rightarrow$\texttt{Architect}$\rightarrow$%
\texttt{Reviewer}$\rightarrow$\texttt{Refiner} pipeline---rather than running
its full development workflow. All agent outputs are evaluated with the same L0--L2 pipeline. All agents run under their default configurations. We use greedy decoding ($temperature=0$) to reduce sampling variance and improve reproducibility; remaining configurations are released in our repository.

\subsection{RQ1: Overall Capability and Input Source}

\begin{table*}[t]
\centering
\caption{Main results on \ourbench{} across the E-R2A and G-R2A subsets, with per-prediction rows pooled before aggregation. SV: syntax-valid rate. NodeCov / GenPrec: reference-side node coverage and generated-node precision. Layer: layer accuracy. GED Acc.: normalized graph-edit-distance accuracy. Orph.: fraction of generated leaf nodes with no incident edges (lower is better). Bold marks the best value per column; underline the best among the four frameworks for a given model.}
\label{tab:rq1-pooled-main-results}
\scriptsize
\resizebox{\textwidth}{!}{%
\begin{tabular}{cl*{12}{r}}
\toprule
 & & \multicolumn{7}{c}{Layer 1: Structural Graph Metrics} & \multicolumn{5}{c}{Layer 2: Judge Scores} \\
\cmidrule(lr){3-9} \cmidrule(lr){10-14}
Model & Framework & SV $\uparrow$ & NodeCov $\uparrow$ & GenPrec $\uparrow$ & Edge F1 $\uparrow$ & GED Acc. $\uparrow$ & Layer $\uparrow$ & Orph. $\downarrow$ & Comp. $\uparrow$ & Faith. $\uparrow$ & Rat. $\uparrow$ & Trace. $\uparrow$ & Read. $\uparrow$ \\
\midrule
\multirow{4}{*}{GPT-5} & Direct & 0.8971 & 0.7945 & 0.5347 & 0.0791 & 38.19 & 0.7496 & 5.88\% & \textbf{\underline{3.87}} & \textbf{\underline{2.72}} & \textbf{\underline{3.39}} & \textbf{\underline{2.57}} & 3.89 \\
 & MetaGPT & \underline{0.9559} & \underline{0.8332} & 0.5112 & 0.1047 & 46.38 & \underline{0.8047} & \textbf{\underline{2.58\%}} & 3.25 & 2.17 & 2.97 & 2.05 & 3.77 \\
 & Mini-SWE & 0.8382 & 0.7696 & \underline{0.6193} & \underline{0.1320} & \underline{49.31} & 0.7785 & 7.13\% & 3.42 & 2.51 & 3.16 & 2.16 & \textbf{\underline{3.96}} \\
 & OpenHands & 0.9265 & 0.7721 & 0.6147 & 0.0909 & 45.49 & 0.7677 & 10.63\% & 3.67 & 2.60 & 3.32 & 2.35 & 3.95 \\
\midrule
\multirow{4}{*}{Claude Sonnet 4.6} & Direct & 0.9706 & \textbf{\underline{0.8490}} & 0.5196 & 0.0804 & 43.52 & 0.8707 & 10.34\% & \underline{3.68} & \underline{2.47} & \underline{3.20} & \underline{2.38} & 3.76 \\
 & MetaGPT & 0.8971 & 0.8201 & 0.5771 & 0.1008 & 48.26 & 0.7887 & \underline{3.20\%} & 3.39 & 2.23 & 3.13 & 2.15 & 3.79 \\
 & Mini-SWE & 0.9118 & 0.7850 & 0.5225 & 0.1243 & 48.84 & 0.7472 & 3.96\% & 3.24 & 2.21 & 2.94 & 2.05 & 3.87 \\
 & OpenHands & \textbf{\underline{1.0000}} & 0.7541 & \underline{0.8949} & \underline{0.1582} & \underline{51.89} & \underline{0.8964} & 13.70\% & 3.60 & 2.41 & 3.14 & 2.23 & \underline{3.91} \\
\midrule
\multirow{4}{*}{DeepSeek V3.2} & Direct & 0.7059 & 0.7882 & 0.4958 & 0.1091 & 44.59 & 0.7628 & 13.00\% & \underline{3.29} & \underline{2.19} & \underline{3.00} & \underline{2.06} & \underline{3.90} \\
 & MetaGPT & \underline{0.7647} & 0.7910 & 0.5969 & \underline{0.1405} & \underline{50.27} & 0.7670 & \underline{6.75\%} & 2.77 & 2.02 & 2.63 & 1.79 & 3.46 \\
 & Mini-SWE & \underline{0.7647} & \underline{0.8437} & 0.5092 & 0.0704 & 45.93 & 0.8297 & 55.70\% & 3.08 & 2.08 & 2.85 & 1.94 & 3.60 \\
 & OpenHands & 0.4853 & 0.8261 & \underline{0.9063} & 0.0987 & 43.61 & \textbf{\underline{0.9259}} & 35.96\% & 3.16 & 2.04 & 2.92 & 2.04 & 3.71 \\
\midrule
\multirow{4}{*}{Qwen3-Coder 480B-A35B} & Direct & 0.5735 & \underline{0.7981} & 0.6105 & 0.0877 & 47.85 & 0.7984 & 10.67\% & \underline{3.08} & \underline{2.26} & \underline{2.95} & \underline{1.95} & \underline{3.95} \\
 & MetaGPT & 0.7206 & 0.7284 & 0.5617 & 0.1057 & 49.12 & 0.7789 & 12.04\% & 2.67 & 1.98 & 2.39 & 1.63 & 3.41 \\
 & Mini-SWE & \underline{0.7794} & 0.6694 & 0.6413 & 0.1060 & 51.76 & 0.7549 & 6.89\% & 2.66 & 2.06 & 2.68 & 1.70 & 3.92 \\
 & OpenHands & 0.5441 & 0.7926 & \textbf{\underline{0.9116}} & \textbf{\underline{0.1758}} & \textbf{\underline{52.82}} & \underline{0.9228} & \underline{4.54\%} & 2.97 & 2.14 & 2.84 & 1.86 & \underline{3.95} \\
\bottomrule
\end{tabular}%
}
\vspace{-5pt}
\end{table*}
\begin{table*}[t]
\centering
\caption{Source-wise performance shift from E-R2A subset to G-R2A subset. NodeCov is reference-side node coverage, and GenPrec is generated-node precision. The delta row reports G-R2A minus E-R2A.}
\label{tab:rq1-compare}
\scriptsize
\begin{tabular}{lrr*{10}{r}}
\toprule
Source & Rows & Valid & SV $\uparrow$ & NodeCov $\uparrow$ & GenPrec $\uparrow$ & Edge F1 $\uparrow$ & GED $\uparrow$ & Layer $\uparrow$ & L2 Avg. $\uparrow$ & Req. Cov. $\uparrow$ & ASR Cov. $\uparrow$ & Unsup. Inf. $\downarrow$ \\
\midrule
E-R2A & 272 & 211/272 & 0.7757 & 0.7682 & 0.6287 & 0.1562 & 50.53 & 0.7528 & 2.82 & 67.24\% & 49.39\% & 32.67\% \\
G-R2A & 816 & 655/816 & 0.8027 & 0.7947 & 0.6079 & 0.0947 & 46.26 & 0.8208 & 2.91 & 69.78\% & 58.13\% & 33.01\% \\
\midrule
G-E & +544 & +444 & +0.0270 & +0.0265 & -0.0209 & -0.0615 & -4.27 & +0.0679 & +0.09 & +2.54\% & +8.74\% & +0.34\% \\
\bottomrule
\end{tabular}
\vspace{-15pt}
\end{table*}
We provide LLMs and agents with full SRSs and request PlantUML architecture views, evaluating their basic ability to derive diagrams from SRS, and we further examine how this ability differs across the two input sources.

Table~\ref{tab:rq1-pooled-main-results} reports the overall results on \ourbench{}. Current LLMs and agentic frameworks generally produce syntactically valid views, but syntax validity remains strongly model- and framework-dependent. Averaged across the four generation settings, closed-source models retain a clear advantage over open-source ones (mean SV around 0.90--0.94 vs.\ 0.65--0.68). Framework choice compounds this: OpenHands reaches perfect SV with Claude Sonnet 4.6 yet is far less stable for the open-source models, dropping below 0.55.

Component recovery is consistently stronger than relation recovery. NodeCov ranges from 0.6694 to 0.8490, so valid outputs usually recover a substantial fraction of the reference components, whereas GenPrec varies far more widely (0.4958--0.9116). OpenHands markedly raises generated-node precision for Claude Sonnet 4.6, DeepSeek V3.2, and Qwen3-Coder, indicating fewer ungrounded components in its valid outputs. Because L1 metrics are conditional on L0-passing outputs, this precision must be read alongside SV: high precision among valid outputs does not by itself imply robust end-to-end generation.

Relation reconstruction is the hardest structural dimension: Edge F1 never exceeds 0.1758 and GED accuracy peaks at 52.82. The strongest structural configuration is Qwen3-Coder with OpenHands, which tops GenPrec, Edge F1, and GED accuracy; however, its low SV and weaker L2 scores qualify this structural lead.

To test whether low Edge F1 reflects a modeling limitation or a property of the task, we manually labeled a stratified random sample of 233 of the 589 missing edges. We find that 65 (27.9\%, Wilson 95\% CI $[22.5\%, 34.0\%]$) are not derivable from the SRS, while the remaining 168 (72.1\%) are at least partially entailed by it; thus part of the gap is an information gap between requirements and the reference architecture. This does not explain the result away: restricting the reference to directly derivable relations raises aggregate Edge F1 only to $\approx\!11.0\%$, still far below node-level alignment. Relation recovery is therefore bounded by both incomplete architectural evidence in the input and limited relational modeling capability.

The semantic (L2) picture diverges from the structural one. GPT-5 with Direct prompting is the strongest semantic configuration, leading completeness, faithfulness, architectural rationality, and traceability; only readability goes instead to GPT-5 with Mini-SWE. The configurations that optimize structural similarity are not those that maximize semantic adequacy: Qwen3-Coder with OpenHands leads Edge F1 and GED accuracy yet trails GPT-5 Direct on every L2 dimension. Topological similarity is thus a useful but insufficient proxy for architecture quality.

Agentic frameworks help selectively rather than uniformly: each improves specific structural or stability metrics for some models (e.g., MetaGPT on GPT-5's coverage and stability), but the gains neither transfer consistently across models nor extend to semantic quality, reshaping the error profile rather than delivering a reliable end-to-end advantage.

We further compare performance across the two input sources by pooling all configurations within each source, as shown in Table~\ref{tab:rq1-compare}. Moving from E-R2A to G-R2A does not uniformly reduce quality. Syntax validity and the average L2 score both rise slightly (0.7757 $\rightarrow$ 0.8027 and 2.82 $\rightarrow$ 2.91), as does requirement coverage (+2.54 percentage points), while ASR coverage rises more clearly (+8.74 percentage points). With fewer components and shallower hierarchies (Table~\ref{tab:benchmark-stats}), G-R2A places less burden on decomposition while still providing enough requirement information for models to produce semantically plausible and requirement-aware views. The degradation is instead concentrated in relation-level structure: Edge F1 drops sharply (0.1562 $\rightarrow$ 0.0947) and GED accuracy decreases (50.53 $\rightarrow$ 46.26), whereas node-level metrics largely hold up---NodeCov even rises slightly (0.7682 $\rightarrow$ 0.7947), GenPrec dips only marginally ($-0.0209$), and layer accuracy rises (0.7528 $\rightarrow$ 0.8208), the last consistent with G-R2A's shallower layer hierarchies. Models thus place components into plausible architectural regions even when they fail to recover the exact reference relations, mirroring the higher relational density of G-R2A reported in Table~\ref{tab:benchmark-stats}.

\begin{tcolorbox}[
    enhanced,
    breakable,
    colback=FindingBgBlue, 
    colframe=FindingFrameBlue, 
    arc=4pt, 
    boxrule=1pt, 
    left=6pt, right=6pt, top=6pt, bottom=6pt,
    drop fuzzy shadow=gray!30
]
\textbf{Finding 1:} Current LLMs and agents identify components and layers far more reliably than they recover relations, and this gap is the dominant limitation across both settings. Moving to projects collected from GitHub degrades relation- and graph-level structure specifically, leaving semantic and evidence-oriented quality largely intact.
\end{tcolorbox}

\subsection{RQ2: View Behavior and Granularity}

We use the manually coded metadata of the G-R2A architecture views to group samples by behavior and granularity, examining whether generation quality varies with the type and abstraction level of the target view. Behavior distinguishes static, dynamic, or mixed views, and granularity indicates a high, medium, or low level of detail.

\begin{table}[t]
\centering
\scriptsize
\caption{L0/L1 evaluation results by manually coded architecture-view categories. Cells report pass rate for L0 and mean values for L1 metrics. NodeCov reports reference-side node coverage, and GenPrec reports generated-node precision.}
\label{tab:rq3-l0-l1-granularity-behavior}
\resizebox{\columnwidth}{!}{%
\begin{tabular}{llrrrrrrr}
\toprule
View factor & Category & L0 Pass & NodeCov & GenPrec & Edge F1 & Layer & GED & Orphan \\
\midrule
Granularity & High-level & 79.1\% & 0.83 & 0.69 & 0.08 & 0.85 & 44.99 & 0.14 \\
Granularity & Medium & 81.2\% & 0.74 & 0.55 & 0.11 & 0.78 & 47.79 & 0.10 \\
Granularity & Low-level & 81.2\% & 0.88 & 0.52 & 0.06 & 0.91 & 44.94 & 0.06 \\
\midrule
Behavior & Static & 82.8\% & 0.78 & 0.62 & 0.09 & 0.82 & 47.88 & 0.11 \\
Behavior & Dynamic & 75.0\% & 0.79 & 0.58 & 0.10 & 0.79 & 43.86 & 0.11 \\
Behavior & Mixed & 83.3\% & 0.90 & 0.63 & 0.10 & 0.91 & 44.96 & 0.11 \\
\bottomrule
\end{tabular}
}%
\vspace{-15pt}
\end{table}

\begin{figure}[htbp]
    \centering
    \setlength{\abovecaptionskip}{0.1cm}
    \begin{subfigure}{\linewidth}
        \centering
        \includegraphics[width=\linewidth]{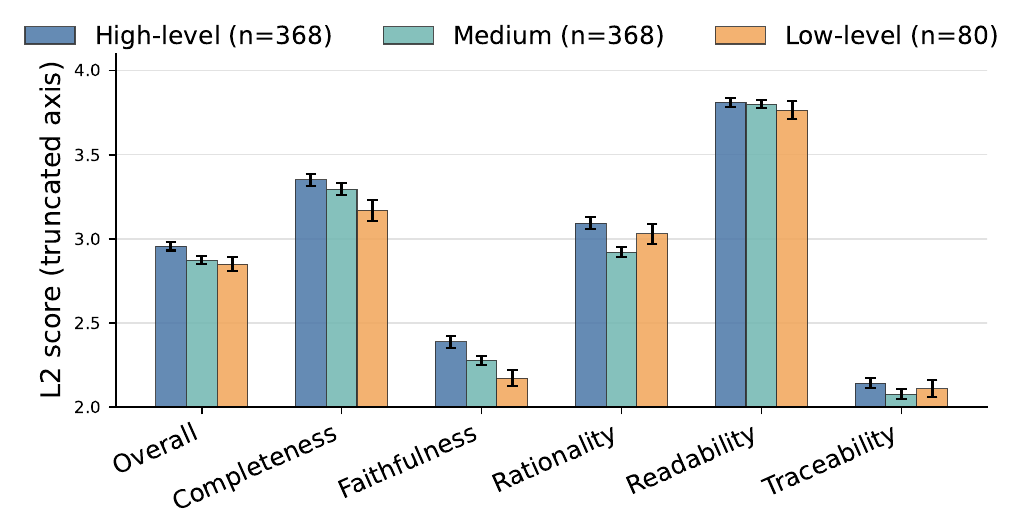}
        \caption{By architectural granularity}
        \label{fig:granularity}
    \end{subfigure}
    \begin{subfigure}{\linewidth}
        \centering
        \includegraphics[width=\linewidth]{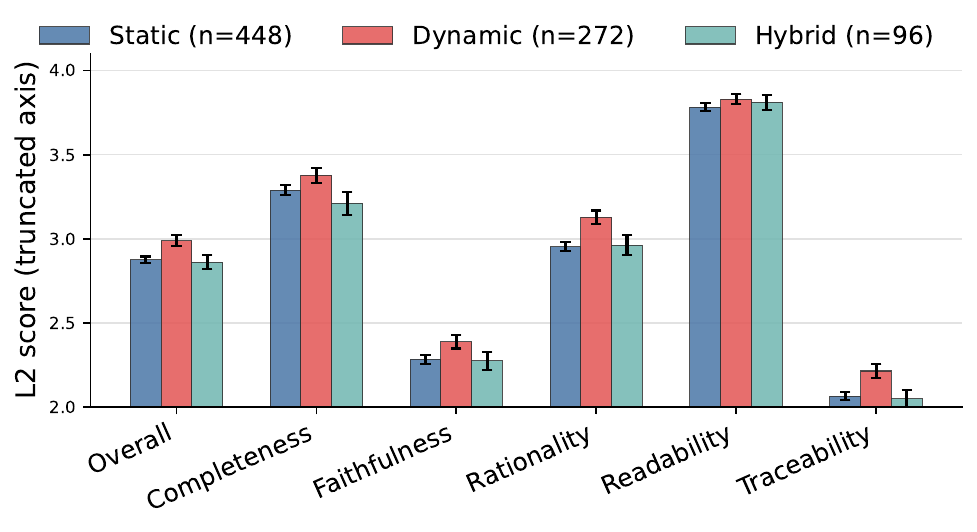}
        \caption{By architectural behavior}
        \label{fig:behavior}
    \end{subfigure}
    \caption{L2 semantic quality scores by architectural granularity (a) and behavior (b).}
    \label{fig:rq2-l2}
    \vspace{-16pt}
\end{figure}

\textbf{Granularity.} As shown in Fig.~\ref{fig:granularity}, high-level views obtain the best semantic scores, leading on overall L2 and on all five L2 dimensions, while medium- and low-level views trail slightly. This suggests repository-derived generation is more reliable at higher abstraction, where the generator need not commit to concrete components and relations that are hard to infer faithfully from distributed artifacts. The L0/L1 metrics show a more mixed pattern: low-level views achieve the best NodeCov, layer accuracy, and orphan ratio, yet their GenPrec stays well below high-level views, so finer-grained targets still encourage unsupported components. Edge F1 remains low across all granularities (0.06--0.11).

\textbf{Behavior.} Behavior shows a similar split. Dynamic views obtain the strongest semantic quality, leading on overall L2 and all five dimensions over static and mixed views, plausibly because behavior-oriented views provide clearer cues about interactions and responsibilities. Structurally, however, mixed views perform best on several L0/L1 indicators and static views achieve the best GED accuracy, so dynamic views lead semantically without dominating the structural metrics.

Across both factors, semantic and structural quality therefore emphasize different aspects of the target view. High-level and dynamic views receive stronger semantic judgments, whereas L0/L1 quality is metric-specific: low-level and mixed views improve component coverage and layer placement, while medium/static categories lead graph-edit similarity. Edge F1 stays low throughout (0.06--0.11 across granularity, 0.09--0.10 across behavior), so the component--relation gap from RQ1 persists across view categories.

\begin{tcolorbox}[
    enhanced,
    colback=FindingBgBlue, 
    colframe=FindingFrameBlue, 
    arc=4pt, 
    boxrule=1pt, 
    left=6pt, right=6pt, top=6pt, bottom=6pt,
    drop fuzzy shadow=gray!30
]
\textbf{Finding 2:} Generation quality is higher for high-level and dynamic views in semantic evaluation, while structural quality varies by metric; across all categories, relation-level accuracy remains the main bottleneck.
\end{tcolorbox}

\subsection{RQ3: Failure Modes}

We analyze recurring R2A failures across the three evaluation layers: rendering and parsing failures at L0, structural mismatches against the reference graph at L1, and semantic and evidence-related failures at L2. We identify them by combining metric outputs, judge explanations, and manual inspection of representative cases.

Applying the L0 validity gate, 866 of 1088 candidates passed and 222 failed, almost all due to ordinary PlantUML syntax or rendering errors rather than environment or dependency issues.

\begin{figure}[htbp]
    \centering
    \setlength{\abovecaptionskip}{0.1cm}
    \includegraphics[width=1.0\linewidth]{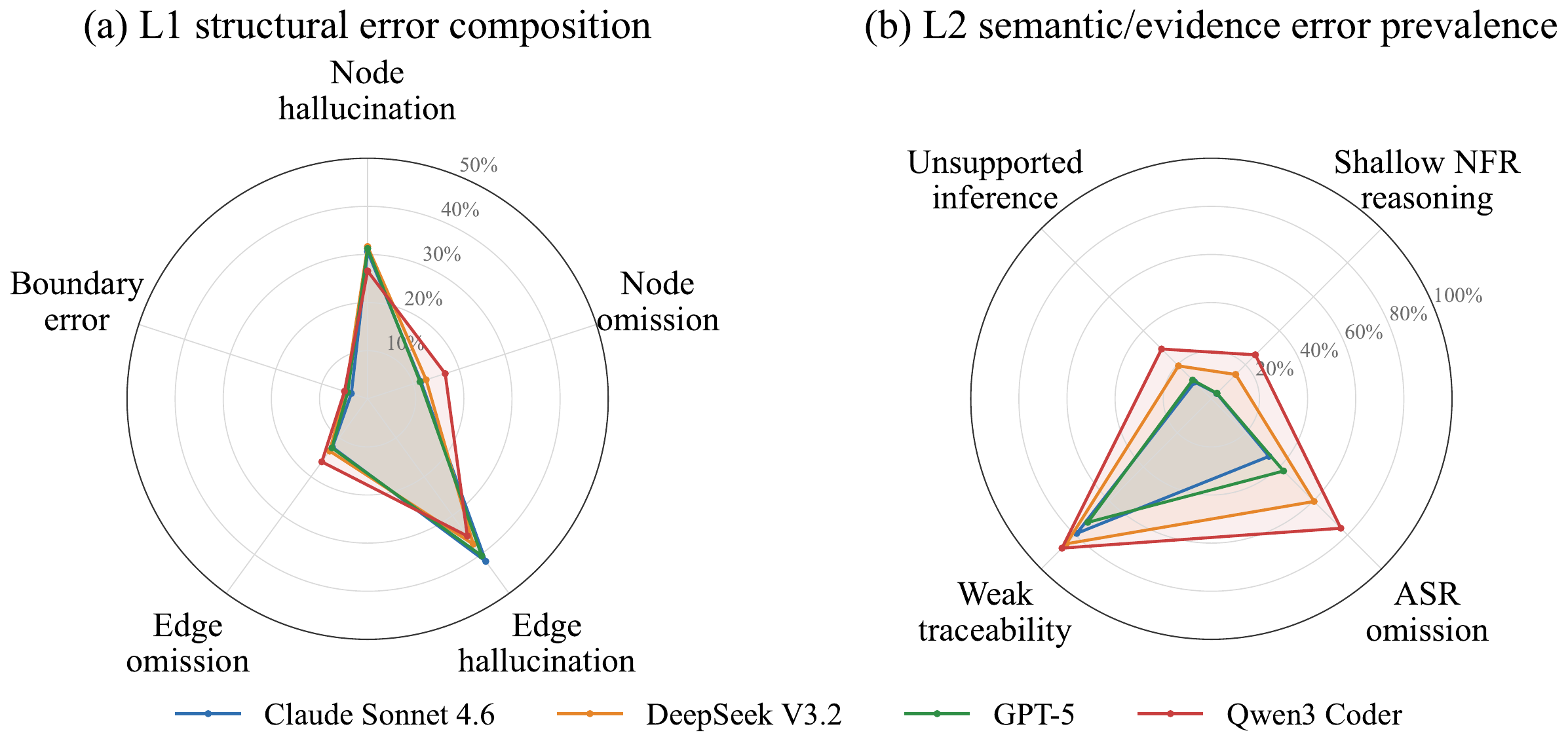}
    \caption{Error analysis across models. (a) Distribution of the five L1 structural error categories; (b) prevalence of the four L2 semantic and evidence-related error categories.}
    \label{fig:error}
    \vspace{-17pt}
\end{figure}

At L1, we decompose structural mismatches into five categories---node and edge omission, node and edge hallucination, and layer error---over the 866 L1-evaluable outputs. Relation-level over-generation dominates: edge hallucination is the largest category (23,820 errors, 39.1\%), followed by node hallucination (18,443, 30.2\%). Edge and node omission account for 8,201 (13.5\%) and 7,859 (12.9\%), while layer errors are far rarer (2,666, 4.4\%). Models thus more often introduce unsupported components and dependencies than misplace matched components across layers. As shown in Fig.~\ref{fig:error}(a), Claude Sonnet 4.6, GPT-5, and DeepSeek V3.2 share a similar over-generation pattern, whereas Qwen3-Coder has relatively higher omission rates.

\vspace{-5pt}

\begin{table}[htbp]
\centering
\scriptsize
\caption{L2 semantic and evidence-related error categories.}
\label{tab:l2-error-categories}
\begin{tabular}{>{\footnotesize}m{0.26\columnwidth}m{0.64\columnwidth}}
\toprule
\textbf{Error Type} & \textbf{Typical Symptom} \\
\midrule
ASR omission & Architecturally significant requirements are missing or not reflected in the view. \\
\midrule
NFR neglect & Quality attributes are stated but not realized as concrete architecture mechanisms. \\
\midrule
Weak traceability & Architecture elements cannot be linked to any requirement or evidence. \\
\midrule
Unsupported claim& Plausible technologies or dependencies are introduced without supporting evidence. \\
\bottomrule
\end{tabular}
\vspace{-5pt}
\end{table}

At L2, we analyze semantic and evidence-related failures. Table~\ref{tab:l2-error-categories} defines four recurring error categories that connect low L2 scores to concrete R2A failure modes. Unlike L1 graph errors, these categories are not mutually exclusive: a single view may, for instance, both omit an ASR and leave its components untraceable. We therefore report, in Fig.~\ref{fig:error}(b), the prevalence rate of each category---the proportion of L2-evaluable outputs in which it appears---rather than mutually exclusive counts.

Across models, the patterns are broadly similar: weak traceability and ASR omission are the most frequent failures, while unsupported inference and NFR neglect are comparatively rare---models fail less by inventing ungrounded elements than by failing to trace and fully cover requirements. Closed-source models perform better overall: Claude Sonnet 4.6 and GPT-5 show lower prevalence across most categories than Qwen3-Coder and DeepSeek V3.2, particularly on ASR omission and traceability. This likely reflects stronger long-context understanding, requirement abstraction, and evidence alignment in closed-source models.

\vspace{-5pt}
\begin{figure}[htbp]
    \centering
    \setlength{\abovecaptionskip}{0.1cm}
    \includegraphics[width=\linewidth]{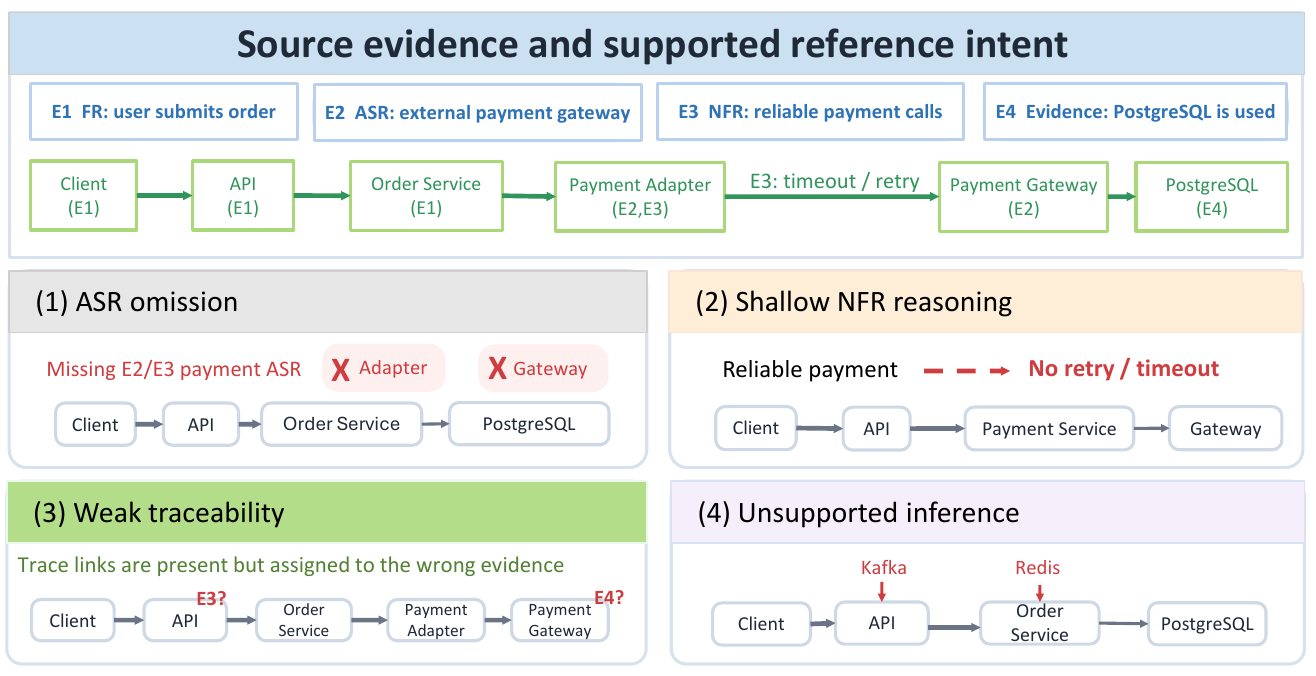}
    \caption{L2 case study}
    \label{fig:l2-case-study}
    \vspace{-18pt}
\end{figure}

To make these categories concrete, Fig.~\ref{fig:l2-case-study} presents an illustrative L2 error case whose reference intent is grounded in four evidence snippets (order submission, an external payment gateway, a reliability requirement on payment calls, and PostgreSQL as the persistence technology). A faithful view should preserve the order-processing path, include the payment adapter and gateway, realize the reliability NFR through concrete mechanisms such as timeout and retry, and keep each element traceable. The lower panels show how a structurally plausible view still fails: omitting the payment ASR, reducing the reliability NFR to a vague label, mislinking evidence, or adding unsupported infrastructure (e.g., Kafka, Redis, a service mesh). This illustrates why L2 targets requirement coverage, NFR realization, traceability, and evidence grounding rather than graph-level similarity alone.

\section{Discussion}

A central lesson from our results is that graph-level agreement and architecture quality come apart. A view may match the reference structurally yet lack evidence support or miss key quality drivers, while two structurally different views can both be reasonable under ambiguous requirements or alternative decompositions. This is why \ourbench{} does not collapse quality into a single score: L1 serves as a reference-based diagnostic that localizes missing or hallucinated components, incorrect layer placement, and erroneous dependencies, while semantic adequacy, traceability, and ASR coverage---which our study finds are where models most often fail---are assessed separately at L2. Progress on R2A therefore cannot be read off diagram validity or topological similarity alone.

\section{Related Work}

\subsection{Software Architecture Design and Architecture Views}

Software architecture design connects requirements, quality attributes, and implementation constraints through explicit structural decisions. Classical architecture-design research emphasizes that architecture is not only a collection of components, but also a set of views, viewpoints, decisions, and tradeoffs that address stakeholder concerns~\cite{hofmeister2007general,nuseibeh2001weaving,IEEE42010}. Architecture-view notations and practices, including C4-style descriptions and quality-attribute-driven design, help architects choose an abstraction level, separate concerns, and reason about system qualities such as modifiability, performance, security, and reliability~\cite{brown2026c4,ISO25010,wojcik2006attribute}. Recent empirical work on open-source architecture views further shows that architecture information in repositories is heterogeneous and often spread across different representation styles~\cite{migliorini2024architectural}. Our work builds on this view-oriented tradition, but studies a different question: whether LLMs and agents can derive a traceable architecture view from requirements and repository evidence, and how the quality of such views can be measured.

\subsection{LLMs for Software Architecture}

LLMs are increasingly being explored for software architecture tasks, including design-decision support, pattern recognition, architecture documentation, and architecture generation from requirements~\cite{schmid2025software,esposito2025generative}. More focused studies have examined whether LLMs can generate architectural design decisions or ADR-like artifacts~\cite{dhar2024can}, transform informal specifications into UML component diagrams~\cite{tagliaferro2025leveraging}, map natural-language requirements to architecture choices with explicit trace links~\cite{helmi2025arlo}, and coordinate multiple LLM agents for C4 architecture modeling~\cite{szczepanik2025collaborative}. These studies indicate that LLMs can assist architecture work but expose recurring limitations: outputs still require expert review, evaluation settings are small or artifact-specific, and generated decisions often lack explicit evidence support. These limitations motivate treating requirements-to-architecture generation as a benchmarked, evidence-grounded task rather than a single prompt.

\subsection{Benchmarks and Evaluation for Architecture Generation}
\label{sec:related-work}
A growing body of LLM-oriented benchmarks targets text-to-model generation, producing a specific artifact type from natural-language input, such as domain models~\cite{chen2023automated,ferrari2024model}, UML class diagrams~\cite{alaswad2026plantucd}, sequence diagrams~\cite{shbita2025mermaidseqbench,ahmed2025mcet}, activity diagrams~\cite{khamsepour2025impact}, and constrained graph models~\cite{chen2025accurate}. While these benchmarks establish useful task formulations and reference oracles, they usually rely on curated, self-contained descriptions that differ from realistic requirement documents, and they target individual diagram types rather than system-level architecture views. Closer to our setting, architecture-specific resources have begun to appear: Migliorini et al. mine architecture views from open-source repositories~\cite{migliorini2024architectural}, recent studies evaluate LLM-based architecture-view generation directly from source-code repositories~\cite{sathvika2026llm}, and ArchBench proposes a platform for benchmarking generative AI on software-architecture tasks~\cite{adnan2026archbench}. However, these resources mainly emphasize existing diagrams, source-code-to-view recovery, or benchmark infrastructure, offering limited support for requirement-grounded inputs and explicit requirement-to-architecture traceability.

Table~\ref{tab:benchmark-capability-comparison} makes this gap concrete, comparing representative datasets and benchmarks along source artifacts, target outputs, reference oracles, traceability support, and layered evaluation. No prior benchmark jointly couples realistic requirements, heterogeneous repository evidence, reference architecture views, and traceable multi-level evaluation---the gap \ourbench{} is designed to fill.

Evaluating generated architecture views is a related challenge, because architecture generation lacks a single executable oracle. Unlike repository-level coding benchmarks such as SWE-bench, where executable patch validation provides the main success signal~\cite{jimenez2023swe}, an architecture view may be syntactically valid yet incomplete, unsupported, or semantically weak, and the same requirements may admit multiple reasonable decompositions. Consequently, text-based similarity metrics are inadequate, while purely rule-based or single-judge LLM checks are either too rigid or vulnerable to bias and hallucination~\cite{tagliaferro2025leveraging,zheng2023judging}. These observations motivate a benchmark whose evaluation exposes multiple complementary dimensions---validity, structural consistency, semantic adequacy, and evidence support---rather than collapsing architecture quality into a single score.

\begin{table}[t]
\centering
\caption{Capability comparison of datasets and benchmarks for model and architecture generation.}
\label{tab:benchmark-capability-comparison}
\tiny
\setlength{\tabcolsep}{3pt}
\renewcommand{\arraystretch}{1.2}
\begin{tabularx}{\columnwidth}{@{}>{\raggedright\arraybackslash}X >{\raggedright\arraybackslash}X *{6}{c}@{}}
\toprule
\textbf{Dataset / Benchmark} & \textbf{Primary source} & \textbf{Req.} & \textbf{Repo} & \textbf{View} & \textbf{Ref.} & \textbf{Trace} & \textbf{Eval} \\
\midrule
UML in GitHub~\cite{robles2017extensive} & GitHub UML artifacts & \xmark & \pmark & \pmark & \xmark & \xmark & \xmark \\
ModelSet~\cite{lopez2022modelset} & GitHub / GenMyModel & \xmark & \xmark & \xmark & \pmark & \xmark & \xmark \\
Ext. EA ModelSet~\cite{glaser2025extended} & Curated ArchiMate & \xmark & \xmark & \pmark & \pmark & \xmark & \xmark \\
Golden UML ModelSet~\cite{verbruggen2025toward} & Curated UML cases & \pmark & \xmark & \xmark & \cmark & \xmark & \pmark \\
Auto. Domain Modeling~\cite{chen2023automated} & Domain descriptions & \cmark & \xmark & \xmark & \cmark & \xmark & \pmark \\
Ferrari et al.~\cite{ferrari2024model} & Requirements docs & \cmark & \xmark & \xmark & \pmark & \xmark & \pmark \\
PLANTUCD~\cite{alaswad2026plantucd} & Requirement--diagram & \cmark & \xmark & \xmark & \cmark & \xmark & \pmark \\
MermaidSeqBench~\cite{shbita2025mermaidseqbench} & Interaction prompts & \cmark & \xmark & \xmark & \cmark & \xmark & \pmark \\
MCeT~\cite{ahmed2025mcet} & Req.\ + sequence & \cmark & \xmark & \xmark & \pmark & \pmark & \cmark \\
LADEX~\cite{khamsepour2025impact} & Process descriptions & \cmark & \xmark & \xmark & \cmark & \pmark & \cmark \\
Graph-model gen.~\cite{chen2025accurate} & Text-to-graph data & \cmark & \xmark & \xmark & \cmark & \xmark & \cmark \\
Arch. views in OSS~\cite{migliorini2024architectural} & GitHub arch. views & \xmark & \pmark & \cmark & \pmark & \xmark & \xmark \\
LLM view gen.~\cite{sathvika2026llm} & Source-code repos & \xmark & \cmark & \cmark & \cmark & \xmark & \pmark \\
ArchBench~\cite{adnan2026archbench} & Aggregated arch. tasks & \pmark & \pmark & \cmark & \pmark & \pmark & \pmark \\
\textbf{R2ABENCH} & \textbf{Edu.\ + GitHub} & \cmark & \cmark & \cmark & \cmark & \cmark & \cmark \\
\bottomrule
\end{tabularx}

\vspace{4pt}

{\scriptsize\raggedright
\cmark\,= primary capability;\quad \pmark\,= partial / task-specific;\quad \xmark\,= not a primary capability.\par
Req = requirement text;\ Repo = repository evidence;\ View = architecture view;\ Ref = reference oracle;\ Trace = trace/evidence links;\ Eval = layered evaluation.\par}
\vspace{-18pt}
\end{table}

\section{Threats to Validity}
\textbf{Internal validity.} Two factors may bias the measured quality. First, our SRS generation, L1 alignment, and L2 judging all rely on GPT-5.4 series models, while GPT-5 is evaluated, raising a potential same-family preference; the L2 judge is calibrated against an independent human review of 272 sampled outputs, which provides a check on any systematic family bias. Second, G-R2A projects come from public GitHub and may appear in pretraining; yet if memorization dominated, G-R2A should score \emph{higher} than E-R2A, whereas its relation-level metrics are lower (Edge F1 $0.1562 \rightarrow 0.0947$; GED $50.53 \rightarrow 46.26$).

\textbf{Construct validity.} We operationalize architecture quality through syntax validity, structural diagnostics, and semantic and evidence-based assessment; these signals are complementary but read as diagnostic indicators rather than a complete measure of design rationale, trade-offs, or all acceptable decompositions.

\textbf{External validity.} \ourbench{} covers 68 projects from educational and GitHub sources, and the SRS---organized from repository documents and evidence---tends to be more complete than early-stage requirements in real development; our findings thus characterize R2A behavior within this scope, and broadening the domains and scale is left to future work.

\section{Conclusion}

This paper presents \ourbench{}, a benchmark and layered evaluation framework for requirements-to-architecture reasoning that pairs educational projects with GitHub projects to span idealized and repository-grounded conditions. Our empirical study shows that current LLM-based systems can often generate syntactically valid and readable architecture views, but relation-level architecture modeling remains a major bottleneck: models identify components more reliably than they recover dependencies. These findings suggest that R2A generation should be evaluated beyond diagram validity or graph similarity, with attention to semantic adequacy, evidence support and view granularity.

\bibliographystyle{IEEEtran}
\bibliography{bibliography}

@misc{iso2018iso,
  title={ISO/IEC/IEEE International Standard-Systems and software engineering--Life cycle processes--Requirements engineering. ISO/IEC/IEEE 29148: 2018 (E)},
  author={ISO/IEC/IEEE},
  year={2018}
}

@inproceedings{migliorini2024architectural,
  title={Architectural views: the state of practice in open-source software projects},
  author={Migliorini, Sofia and Verdecchia, Roberto and Malavolta, Ivano and Lago, Patricia and Vicario, Enrico},
  booktitle={European Conference on Software Architecture},
  pages={396--415},
  year={2024},
  organization={Springer}
}

@book{robertson2009probabilistic,
  title={The probabilistic relevance framework: BM25 and beyond},
  author={Robertson, Stephen and Zaragoza, Hugo},
  volume={4},
  year={2009},
  publisher={Now Publishers Inc}
}

@article{salton1975vector,
  title={A vector space model for automatic indexing},
  author={Salton, Gerard and Wong, Anita and Yang, Chung-Shu},
  journal={Communications of the ACM},
  volume={18},
  number={11},
  pages={613--620},
  year={1975},
  publisher={ACM New York, NY, USA}
}

@inproceedings{bajracharya2006sourcerer,
  title={Sourcerer: a search engine for open source code supporting structure-based search},
  author={Bajracharya, Sushil and Ngo, Trung and Linstead, Erik and Dou, Yimeng and Rigor, Paul and Baldi, Pierre and Lopes, Cristina},
  booktitle={Companion to the 21st ACM SIGPLAN symposium on Object-oriented programming systems, languages, and applications},
  pages={681--682},
  year={2006}
}

@inproceedings{cormack2009reciprocal,
  title={Reciprocal rank fusion outperforms condorcet and individual rank learning methods},
  author={Cormack, Gordon V and Clarke, Charles LA and Buettcher, Stefan},
  booktitle={Proceedings of the 32nd international ACM SIGIR conference on Research and development in information retrieval},
  pages={758--759},
  year={2009}
}

@article{hofmeister2007general,
  title={A general model of software architecture design derived from five industrial approaches},
  author={Hofmeister, Christine and Kruchten, Philippe and Nord, Robert L and Obbink, Henk and Ran, Alexander and America, Pierre},
  journal={Journal of Systems and Software},
  volume={80},
  number={1},
  pages={106--126},
  year={2007},
  publisher={Elsevier}
}

@article{nuseibeh2001weaving,
  title={Weaving together requirements and architectures},
  author={Nuseibeh, Bashar},
  journal={Computer},
  volume={34},
  number={3},
  pages={115--119},
  year={2001},
  publisher={IEEE}
}

@book{brown2026c4,
  title={The C4 Model: Visualizing Software Architecture},
  author={Brown, Simon},
  year={2026},
  publisher={" O'Reilly Media, Inc."}
}

@ARTICLE{IEEE42010,
  author={},
  journal={ISO/IEC/IEEE 42010:2022(E)}, 
  title={IEEE/ISO/IEC International Standard for Software, systems and enterprise--Architecture description}, 
  year={2022},
  volume={},
  number={},
  pages={1-74},
  doi={10.1109/IEEESTD.2022.9938446}}

@misc{ISO25010,
  author = {{ISO/IEC}},
  title = {{ISO/IEC 25010:2023, Systems and software engineering -- Systems and software Quality Requirements and Evaluation (SQuaRE) -- Product quality model}},
  year = {2023},
  howpublished = {\url{https://www.iso.org/standard/78176.html}},
  note = {Accessed: 2026-06-21}
}

@techreport{wojcik2006attribute,
  title={Attribute-driven design (ADD), version 2.0},
  author={Wojcik, Rob and Bachmann, Felix and Bass, Len and Clements, Paul and Merson, Paulo and Nord, Robert and Wood, Bill},
  institution = {Software Engineering Institute, Carnegie Mellon University},
  year={2006}
}

@article{chen2021evaluating,
  title={Evaluating large language models trained on code},
  author={Chen, Mark and Tworek, Jerry and Jun, Heewoo and Yuan, Qiming and Pinto, Henrique Ponde De Oliveira and Kaplan, Jared and Edwards, Harri and Burda, Yuri and Joseph, Nicholas and Brockman, Greg and others},
  journal={arXiv preprint arXiv:2107.03374},
  year={2021}
}

@article{zheng2023judging,
  title={Judging llm-as-a-judge with mt-bench and chatbot arena},
  author={Zheng, Lianmin and Chiang, Wei-Lin and Sheng, Ying and Zhuang, Siyuan and Wu, Zhanghao and Zhuang, Yonghao and Lin, Zi and Li, Zhuohan and Li, Dacheng and Xing, Eric and others},
  journal={Advances in neural information processing systems},
  volume={36},
  pages={46595--46623},
  year={2023}
}

@inproceedings{medvidovic2010software,
  title={Software architecture: foundations, theory, and practice},
  author={Medvidovic, Nenad and Taylor, Richard N},
  booktitle={Proceedings of the 32nd ACM/IEEE International Conference on Software Engineering-Volume 2},
  pages={471--472},
  year={2010}
}

@book{rozanski2012software,
  title={Software systems architecture: working with stakeholders using viewpoints and perspectives},
  author={Rozanski, Nick and Woods, Eoin},
  year={2012},
  publisher={Addison-Wesley}
}

@inproceedings{clements2003documenting,
  title={Documenting software architectures: views and beyond},
  author={Clements, Paul and Garlan, David and Little, Reed and Nord, Robert and Stafford, Judith},
  booktitle={25th International Conference on Software Engineering, 2003. Proceedings.},
  pages={740--741},
  year={2003},
  organization={IEEE}
}

@inproceedings{tekinerdogan2011defining,
  title={Defining architectural viewpoints for quality concerns},
  author={Tekinerdogan, Bedir and S{\"o}zer, Hasan},
  booktitle={European Conference on Software Architecture},
  pages={26--34},
  year={2011},
  organization={Springer}
}

@article{esposito2025generative,
  title={Generative AI for software architecture. Applications, challenges, and future directions},
  author={Esposito, Matteo and Li, Xiaozhou and Moreschini, Sergio and Ahmad, Noman and Cerny, Tomas and Vaidhyanathan, Karthik and Lenarduzzi, Valentina and Taibi, Davide},
  journal={Journal of Systems and Software},
  pages={112607},
  year={2025},
  publisher={Elsevier}
}

@article{sathvika2026llm,
  title={LLM-based Automated Architecture View Generation: Where Are We Now?},
  author={Sathvika, Miryala and Dhar, Rudra and Vaidhyanathan, Karthik},
  journal={arXiv preprint arXiv:2603.21178},
  year={2026}
}

@article{fleiss1971measuring,
  title={Measuring nominal scale agreement among many raters.},
  author={Fleiss, Joseph L},
  journal={Psychological bulletin},
  volume={76},
  number={5},
  pages={378},
  year={1971},
  publisher={American Psychological Association}
}

@inproceedings{hong2024metagpt,
      title={Meta{GPT}: Meta Programming for A Multi-Agent Collaborative Framework},
      author={Sirui Hong and Mingchen Zhuge and Jonathan Chen and Xiawu Zheng and Yuheng Cheng and Jinlin Wang and Ceyao Zhang and Zili Wang and Steven Ka Shing Yau and Zijuan Lin and Liyang Zhou and Chenyu Ran and Lingfeng Xiao and Chenglin Wu and J{\"u}rgen Schmidhuber},
      booktitle={The Twelfth International Conference on Learning Representations},
      year={2024},
      url={https://openreview.net/forum?id=VtmBAGCN7o}
}

@misc{openhands,
      title={{OpenHands: An Open Platform for AI Software Developers as Generalist Agents}},
      author={Xingyao Wang and Boxuan Li and Yufan Song and Frank F. Xu and Xiangru Tang and Mingchen Zhuge and Jiayi Pan and Yueqi Song and Bowen Li and Jaskirat Singh and Hoang H. Tran and Fuqiang Li and Ren Ma and Mingzhang Zheng and Bill Qian and Yanjun Shao and Niklas Muennighoff and Yizhe Zhang and Binyuan Hui and Junyang Lin and Robert Brennan and Hao Peng and Heng Ji and Graham Neubig},
      year={2024},
      eprint={2407.16741},
      archivePrefix={arXiv},
      primaryClass={cs.SE},
      url={https://arxiv.org/abs/2407.16741},
}

@inproceedings{yang2024sweagent,
  title={{SWE}-agent: Agent-Computer Interfaces Enable Automated Software Engineering},
  author={John Yang and Carlos E Jimenez and Alexander Wettig and Kilian Lieret and Shunyu Yao and Karthik R Narasimhan and Ofir Press},
  booktitle={The Thirty-eighth Annual Conference on Neural Information Processing Systems},
  year={2024},
  url={https://arxiv.org/abs/2405.15793}
}

@article{schmid2025software,
  title={Software architecture meets llms: A systematic literature review},
  author={Schmid, Larissa and Hey, Tobias and Armbruster, Martin and Corallo, Sophie and Fuch{\ss}, Dominik and Keim, Jan and Liu, Haoyu and Koziolek, Anne},
  journal={arXiv preprint arXiv:2505.16697},
  year={2025}
}

@inproceedings{dhar2024can,
  title={Can llms generate architectural design decisions?-an exploratory empirical study},
  author={Dhar, Rudra and Vaidhyanathan, Karthik and Varma, Vasudeva},
  booktitle={2024 IEEE 21st International Conference on Software Architecture (ICSA)},
  pages={79--89},
  year={2024},
  organization={IEEE}
}

@inproceedings{tagliaferro2025leveraging,
  title={Leveraging llms to automate software architecture design from informal specifications},
  author={Tagliaferro, Alberto and Corbo, Simone and Guindani, Bruno},
  booktitle={2025 IEEE 22nd International Conference on Software Architecture Companion (ICSA-C)},
  pages={291--299},
  year={2025},
  organization={IEEE}
}

@article{helmi2025arlo,
  title={ARLO: A Tailorable Approach for Transforming Natural Language Software Requirements into Architecture using LLMs},
  author={Helmi, Tooraj},
  journal={arXiv preprint arXiv:2504.06143},
  year={2025}
}

@article{szczepanik2025collaborative,
  title={Collaborative LLM Agents for C4 Software Architecture Design Automation},
  author={Szczepanik, Kamil and Chudziak, Jaros{\'L} and others},
  journal={arXiv preprint arXiv:2510.22787},
  year={2025}
}

@inproceedings{robles2017extensive,
  title={An extensive dataset of UML models in GitHub},
  author={Robles, Gregorio and Ho-Quang, Truong and Hebig, Regina and Chaudron, Michel RV and Fernandez, Miguel Angel},
  booktitle={2017 IEEE/ACM 14th International Conference on Mining Software Repositories (MSR)},
  pages={519--522},
  year={2017},
  organization={IEEE}
}

@article{lopez2022modelset,
  title={ModelSet: a dataset for machine learning in model-driven engineering},
  author={L{\'o}pez, Jos{\'e} Antonio Hern{\'a}ndez and Canovas Izquierdo, Javier Luis and Cuadrado, Jes{\'u}s S{\'a}nchez},
  journal={Software and Systems Modeling},
  volume={21},
  number={3},
  pages={967--986},
  year={2022},
  publisher={Springer}
}

@article{glaser2025extended,
  title={The extended ea modelset—a fair dataset for researching and reasoning enterprise architecture modeling practices},
  author={Glaser, Philipp-Lorenz and Sallinger, Emanuel and Bork, Dominik},
  journal={Software and Systems Modeling},
  pages={1--19},
  year={2025},
  publisher={Springer}
}

@inproceedings{chen2023automated,
  title={Automated domain modeling with large language models: A comparative study},
  author={Chen, Kua and Yang, Yujing and Chen, Boqi and L{\'o}pez, Jos{\'e} Antonio Hern{\'a}ndez and Mussbacher, Gunter and Varr{\'o}, D{\'a}niel},
  booktitle={2023 ACM/IEEE 26th International Conference on Model Driven Engineering Languages and Systems (MODELS)},
  pages={162--172},
  year={2023},
  organization={IEEE}
}

@inproceedings{ferrari2024model,
  title={Model generation with LLMs: From requirements to UML sequence diagrams},
  author={Ferrari, Alessio and Abualhaija, Sallam and Arora, Chetan},
  booktitle={2024 IEEE 32nd International Requirements Engineering Conference Workshops (REW)},
  pages={291--300},
  year={2024},
  organization={IEEE}
}

@article{alaswad2026plantucd,
  title={PLANTUCD: A DATASET OF SOFTWARE REQUIREMENTS AND CORRESPONDING PLANTUML-BASED CLASS DIAGRAMS},
  author={Alaswad, Feisal and E, Poovammal and Aljaddouh, Batoul},
  year={2026},
  publisher={TechRxiv}
}

@article{shbita2025mermaidseqbench,
  title={MermaidSeqBench: An Evaluation Benchmark for LLM-to-Mermaid Sequence Diagram Generation},
  author={Shbita, Basel and Ahmed, Farhan and DeLuca, Chad},
  journal={arXiv preprint arXiv:2511.14967},
  year={2025}
}

@inproceedings{ahmed2025mcet,
  title={MCeT: Behavioral Model Correctness Evaluation using Large Language Models},
  author={Ahmed, Khaled and Song, Jialing and Chen, Boqi and Wei, Ou and Zheng, Bingzhou},
  booktitle={2025 ACM/IEEE 28th International Conference on Model Driven Engineering Languages and Systems (MODELS)},
  pages={84--95},
  year={2025},
  organization={IEEE}
}

@article{khamsepour2025impact,
  title={The Impact of Critique on LLM-Based Model Generation from Natural Language: The Case of Activity Diagrams},
  author={Khamsepour, Parham and Cole, Mark and Ashraf, Ish and Tan, DaYuan and Puri, Sandeep and Sabetzadeh, Mehrdad and Nejati, Shiva},
  journal={arXiv preprint arXiv:2509.03463},
  year={2025}
}

@inproceedings{chen2025accurate,
  title={Accurate and Consistent Graph Model Generation from Text with Large Language Models},
  author={Chen, Boqi and Wei, Ou and Zheng, Bingzhou and Mussbacher, Gunter},
  booktitle={2025 ACM/IEEE 28th International Conference on Model Driven Engineering Languages and Systems (MODELS)},
  pages={130--141},
  year={2025},
  organization={IEEE}
}

@article{adnan2026archbench,
  title={ArchBench: Benchmarking Generative-AI for Software Architecture Tasks},
  author={Adnan, Bassam and Gupta, Aviral and Akshathala, Sreemaee and Vaidhyanathan, Karthik},
  journal={arXiv preprint arXiv:2603.17833},
  year={2026}
}

@article{jimenez2023swe,
  title={Swe-bench: Can language models resolve real-world github issues?, 2024},
  author={Jimenez, Carlos E and Yang, John and Wettig, Alexander and Yao, Shunyu and Pei, Kexin and Press, Ofir and Narasimhan, Karthik},
  journal={URL https://arxiv. org/abs/2310.06770},
  volume={7},
  year={2023}
}

@inproceedings{papineni2002bleu,
  title={Bleu: a method for automatic evaluation of machine translation},
  author={Papineni, Kishore and Roukos, Salim and Ward, Todd and Zhu, Wei-Jing},
  booktitle={Proceedings of the 40th annual meeting of the Association for Computational Linguistics},
  pages={311--318},
  year={2002}
}

@inproceedings{verbruggen2025toward,
  title={Toward a Community-Curated Golden Dataset of UML Models},
  author={Verbruggen, Charlotte and Netz, Lukas and Glaser, Philipp-Lorenz and Scholz, Marion and Huemer, Christian and Calamo, Marco and Rumpe, Bernhard and Snoeck, Monique and Bork, Dominik},
  booktitle={2025 ACM/IEEE 28th International Conference on Model Driven Engineering Languages and Systems Companion (MODELS-C)},
  pages={43--50},
  year={2025},
  organization={IEEE}
}

\end{document}